\documentclass[manuscript]{acmart}
\renewcommand\footnotetextcopyrightpermission[1]{} 
\usepackage{booktabs} 
\usepackage{multirow}
\usepackage{graphicx}

\usepackage{amsmath}
\usepackage{amsfonts}
\usepackage{graphicx}
\usepackage{wrapfig}

\usepackage[ruled,vlined]{algorithm2e}
\usepackage{algpseudocode}

\newcommand{\squishlist}{
 \begin{list}{$\bullet$}
  { \setlength{\itemsep}{0pt}
     \setlength{\parsep}{3pt}
     \setlength{\topsep}{3pt}
     \setlength{\partopsep}{0pt}
     \setlength{\leftmargin}{1.5em}
     \setlength{\labelwidth}{1em}
     \setlength{\labelsep}{0.5em} } }

\newcommand{\squishlisttwo}{
 \begin{list}{$\bullet$}
  { \setlength{\itemsep}{0pt}
     \setlength{\parsep}{0pt}
    \setlength{\topsep}{0pt}
    \setlength{\partopsep}{0pt}
\setlength{\leftmargin}{2em}
\setlength{\labelwidth}{1.5em}
\setlength{\labelsep}{0.5em} } }

\newcommand{\squishend}{
\end{list}  }

\setcopyright{rightsretained}

\acmDOI{10.475/123_4}

\acmISBN{123-4567-24-567/08/06}

\acmConference[WOODSTOCK'97]{ACM Woodstock conference}{July 1997}{El Paso, Texas USA} 
\acmYear{1997}
\copyrightyear{2016}

\acmArticle{4}
\acmPrice{15.00}

\begin{document}
\title{Exploring Heterogeneous Metadata for Video Recommendation with Two-tower Model}

\author{Jianling Wang}
\affiliation{%
	\institution{Texas A\&M University}
}
\email{jlwang@tamu.edu}

\author{Ainur Yessenalina}
\affiliation{%
	\institution{Amazon Inc.}
}
\email{yessenal@amazon.com}

\author{Alireza Roshan-Ghias}
\affiliation{%
	\institution{Amazon Inc.}
}
\email{ghiasali@amazon.com}

\begin{abstract}
Online video services acquire new content on a daily basis to increase engagement, and improve the user experience. Traditional recommender systems solely rely on watch history, delaying the recommendation of newly added titles to the right customer. However, one can use the metadata information of a cold-start title to bootstrap the personalization. In this work, we propose to adopt a two-tower model, in which one tower is to learn the user representation based on their watch history, and the other tower is to learn the effective representations for titles using metadata. The contribution of this work can be summarized as: (1) we show the feasibility of using two-tower model for recommendations and conduct a series of offline experiments to show its performance for cold-start titles; (2) we explore different types of metadata (categorical features, text description, cover-art image) and an attention layer to fuse them; (3) with our Amazon proprietary data, we show that the attention layer can assign weights adaptively to different metadata with improved recommendation for warm- and cold-start items.
\end{abstract}

%



\maketitle

\section{Introduction}


Online video services like Amazon Prime Video add cold-start contents, i.e. titles that do not have any watch history, on a regular basis to their catalog. Newly added titles are typically the most interesting for users due to the novelty factor. In fact, many users scroll the page endlessly to find new titles that pique their interest. Thus, it is important to match these highly desired titles to the right users as soon as possible. 

\begin{wrapfigure}{R}{0.55\textwidth}
	\includegraphics[width=0.99\linewidth]{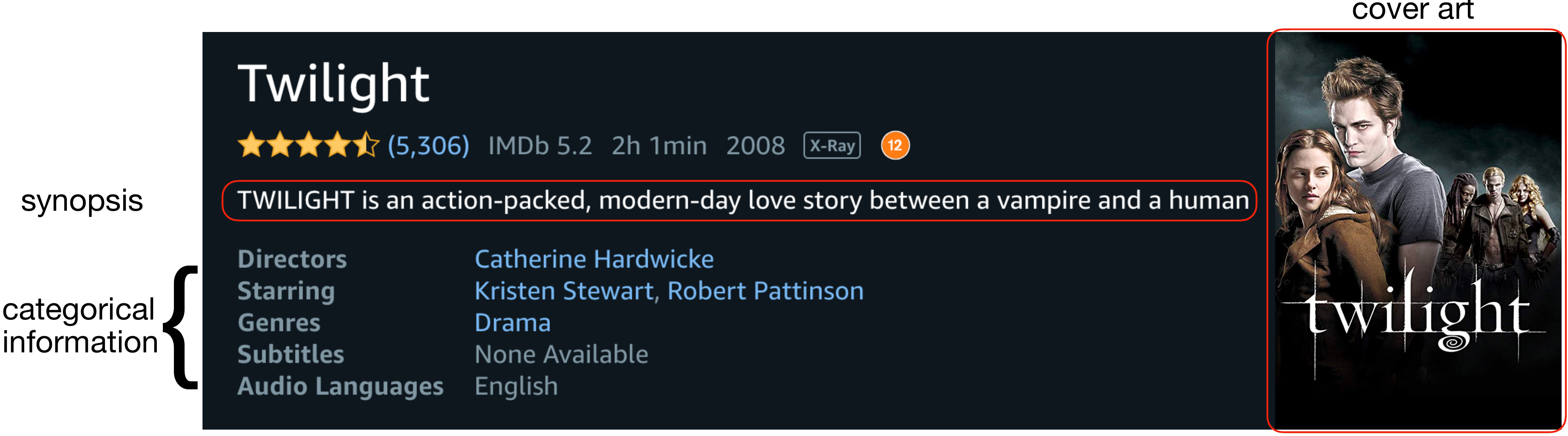}
	\caption{The detail page for a video in Amazon Prime Video. There are different types of metadata for each video, including categorical features, synopsis and cover art.}
	\label{fig:detail}
\end{wrapfigure}

Recommender systems based on watch history such as neural network-based architectures \cite{covington2016deep, Rybakov2018TheEO} or matrix factorization techniques \cite{he2017neural, koren2009matrix} can only generate recommendations for a set of items seen during the model training. This hinders the recommendation to work effectively typically for many days until enough watches have happened. Cold-start titles typically come with different types of metadata, including categorical features (i.e., genre, cast, release time, etc), synopses and cover art images. As in Figure \ref{fig:detail}, metadata always shows up on the content page to characterize the title and it helps users decide whether they are interested in a title or not. In this work, we postulate that we can utilize the metadata of the cold-start titles and match the right title to the right user, based on their previous watch history. Meanwhile, the metadata can work as supplementary information and help to infer users' preferences on the warm-start titles.

In this work, we adopt the two-tower architecture for recommendation of videos with heterogeneous metadata. The item and user representations are learned from two different neural towers, which can be used to infer the user-item preference with dot product operation. To support heterogeneous metadata, we develop an attention fusion layer to combine metadata from different modalities. In this work, we explore the use of three different metadata in Amazon Prime Video: categorical features (i.e., genres, actors and directors and release time), synopsis features, and cover art features. With the experiments, we show how different metadata contribute to the title representation learning and also contribute to both warm- and cold-start item recommendation. 




\section{Related Work}  

Context-aware recommender systems improve recommendations by incorporating contextual information of the user's decision into the recommendation process \cite{gongcontextual, adomavicius2005incorporating,wang2020next}.  Previous research in video recommendation benefits a lot from different types of contextual information. In an early work, \cite{davidson2010youtube} tried to obtain the candidate videos based on co-visitation counts and ranks them utilizing the rule-based signals. To obtain informative user representations for candidate generation, the authors of \cite{covington2016deep} concatenated heterogeneous user features and input them into a deep neural network. To enable the recommendation for video with limited historic feedback, different content-based video recommendation framework is proposed recently. For example, \cite{lee2017large} embedded all the videos relying on their raw video and audio content, and recommended the similar videos located closely on an embedding space. They concatenated heterogeneous user features and generate user embedding with a deep neural network. In our work, we are exploring the impacts of heterogeneous metadata with the two-tower model to conduct efficient user modeling and video metadata modeling at the same time.

Besides modeling user preferences on warm-start videos, we also aim to explore the feasibility of recommending the newly-available cold-start videos. In fact, cold-start recommendation has been explored via mapping between metadata and the well-trained embedding \cite{gantner2010learning} or training strategies like Dropout \cite{volkovs2017dropoutnet} and meta-learning \cite{lee2019melu,wang2021sequential}. These methods usually work on one particular type of metadata. In this work, we are focusing on exploring the feasibility of utilizing heterogeneous metadata in characterizing videos for cold-start recommendation.
 
The two-tower structure we have adopted in this work has been applied in the industry to predict the relevance between a query and the candidate items (e.g., news, apps and documents) \cite{huang2013learning, elkahky2015multi, wangimproving, Rybakov2018TheEO}, in which one tower is used to model the query and the other one is for the candidate item. These works show that the two-tower model can be easily modified and extended for multi-view learning, for cross-domain transfer learning or for supporting different negative sampling strategies. There are also previous works utilizing the two-tower model for user and item representation learning in recommendation systems \cite{yang2020mixed}. However, those works usually rely on only one specific type of metadata or directly concatenate various types of metadata. We adopt the two-tower model for user-video preference prediction task and explore the design of the item tower with different item metadata. Ultimately, the model can generate predictions on both warm and cold-start items.

\section{Two-tower Model}
\subsection{Architecture}
As the name implies, the model has two components: user and item towers, each producing the corresponding embeddings, culminating in a dot product between the two and passing through a sigmoid activation function.

\smallskip
\noindent\textbf{User Tower.} Users are represented by their watch histories in the training time period and some additional user-level features such as their country. We are adopting a user encoder architecture previously tested in production. Due to the focus on item encoder in this work, we did not experiment with different types of user encoder and have its architecture fixed. With the user input, the watch history is firstly encoded with a position-aware attention layer to extract the sequential patterns. The user-level features are directly embedded with an MLP layer. Then the encoder concatenates watch history embeddings and user feature embeddings, passing them through the three residual blocks \cite{he2016deep}, which is introduced to assist in training a deeper architecture and thus help to achieve better performance.



\smallskip
\noindent\textbf{Item Tower.} For each item appearing in the training period, we can use so-called {\bf ID feature} that is in essence the one-hot encoding for all known items. The embedding layer of the ID feature is uniquely linked to the ID of the video and is updated during the training process. Meanwhile, each item also has other metadata available  (genres, actors, directors, etc; synopsis, cover art image), which can be used to generate a dense item representation. More details are provided in the following subsections.

\smallskip
\noindent\textbf{Preference Prediction.}  Given a user-item pair $(u, i)$, we can generate the user embedding $\textbf{u}$ and item embedding $\textbf{i}$ with the corresponding towers. Then following the idea of matrix factorization \cite{koren2009matrix}, we can use the dot product $\textbf{u} \cdot \textbf{i}$ to approximate $u$'s preference on item $i$. Note that we also apply the Sigmoid function $\sigma(\cdot)$ on the dot product $\textbf{u} \cdot \textbf{i}$ as activation. In the training process, we use cross-entropy to calculate the loss. In the prediction process, we will predict users' preference scores on the set of videos and recommend top scoring titles.

\begin{figure}
	\centering
	\includegraphics[width=0.85\textwidth]{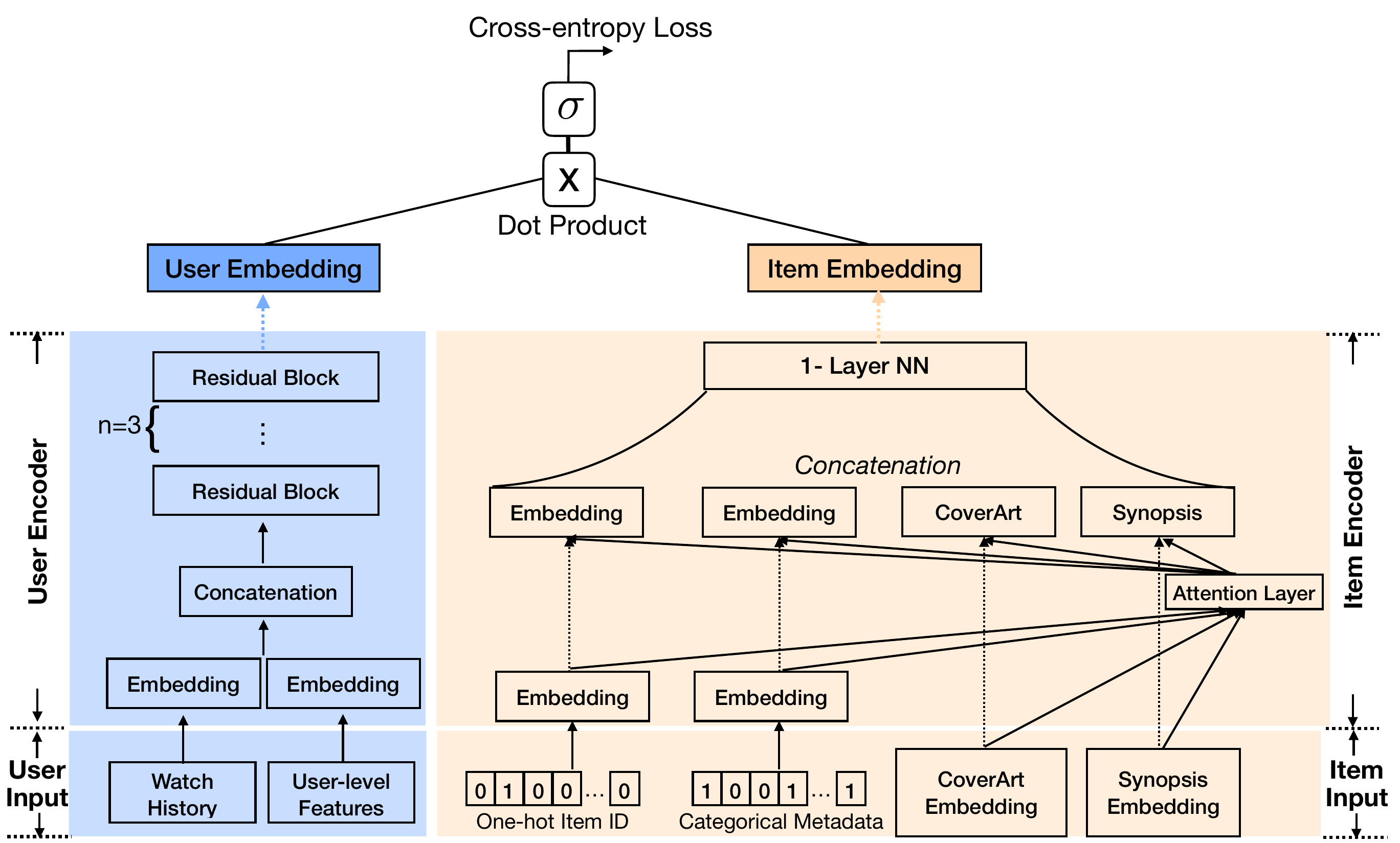}
	\caption{The proposed two-tower model with heterogeneous metadata.}%
	\label{fig:model}%
\end{figure}


\subsection{Metadata}
\label{sec:meta}
In the following section, we will elaborate on three types of metadata used in the item tower: categorical features, synopsis and cover art. 

\smallskip
\noindent\textbf{Categorical features} is a subset of metadata listed on the product page to characterize the videos. They include categorical features represented with a high-dimensional binary vector.  
\begin{itemize}
\item \textbf{Genre} is used to indicate the topic of a title as displayed on the detail page, which covers 27 tags  and 227 subgenre tags indicating more fine-grained topics. We use a binary vector with 227+ 27 dimensions to represent this feature. 
\item \textbf{Actors/Directors} are important factors indicating the relevance of the title to a user. To avoid the long-tail issue, only the top-2000 actors and directors are taken into consideration. 
\item \textbf{Maturity Rating} containing 17 different levels, indicating the maturity level of a title.
 \item \textbf{Country of Origin} containing 159 different values, indicating the country/region title is created in.
\item \textbf{Release Year} indicating the time the title was released. 
\item \textbf{Acquisition Date} indicating when the title became available in Prime Video. We set the granularity to be 1-month to convert the release dates into categorical values.
\item \textbf{Popularity} is based on the number of views in a certain period. We use two different values, one is based on the 2-year watching history and the other one is based on the watching history of the most recent 60 days. We use the total number of views to normalize the values and use the log function to smooth the values. We apply uniform discretization transformation on the resulting values to  covert them into categorical type of input.
\end{itemize}
By concatenating the vectors generated from each of the features above, we obtain a high dimensional vector for each of the title.

\smallskip
\noindent\textbf{Synopsis} is the short text (usually a few sentences) displayed on the detail page which summarizes a title. It can help users to make a decision on whether to watch a title or not. We encode the synopsis using TF-IDF-weighted sum of the word2vec \cite{mikolov2013distributed, mikolov2013efficient} embeddings of each word. 


\smallskip
\noindent\textbf{Cover Art} is the image displayed on the detail page to give a visual for the movie, which can also hint on the content and style of the title. In order to obtain an embedding for the cover art, we use a pre-trained 34-layer ResNet trained on ImageNet data \cite{he2016deep}, and extract the activations before the last layer.

\smallskip
\noindent\textbf{ID} is used to represent titles present during the training. Note that we do not have IDs for cold-start users during training. During the evaluation process, we set the ID embedding to zero for cold-start titles.

\subsection{Attention Layer for Fusion}
After obtaining the ID, categorical, synopsis and cover art embeddings, we concatenate them all together. However, different components can have different influences on the title representations. For example, some titles have informative cover arts but the others may not. Thus we need to fuse the metadata from different modalities considering their importance. We adopt the attention mechanism \cite{vaswani2017attention, wang2020user} to calculate the weights for each component. Let $\alpha_t^m$ denote the attention weight for metadata type $m$ of video $t$ and M includes all the types of metadata that the model considering as input for the item tower. Then we will have the pre-score $ O_t^m$:
\begin{displaymath} 
\alpha_t^m = \frac{\exp O_t^m}{\sum_{k \in M}\exp O_t^k},  \quad O_t^{\textit{m}} = \textbf{z}^T \cdot \tanh (\textbf{P}\textbf{h}_t^{\textit{m}} + \textbf{b})
\end{displaymath}
where $\textbf{h}_t^{\textit{m}}$ represents the embedding for metadata type $m$ of title $t$, \textbf{P} is the weight matrix, \textbf{z} is a transform vector and \textbf{b} is the bias vector. By applying softmax operation on the pre-scores, we can obtain the attention weights of different components which can sum up to be one. After we obtaining the weights, they will be multiplied with the metadata before the concatenation. In the end, a 1-layer perceptron is used to convert the fused embedding to have the same size of the user representation. Given that cold-start items haven't shown up in the training, the ID embedding will be a zero vector. Thus the item representation is purely based on its metadata.

\section{Experiment}
In this section, we will first elaborate the experiment setup and then explore the feasibility of utilizing heterogeous metadata for both warm- and cold-start items in video recommendation with experiments on Amazon Prime Video streaming history.  

\begin{wrapfigure}{R}{0.55\textwidth}
	\centering
		\includegraphics[width=0.5\textwidth]{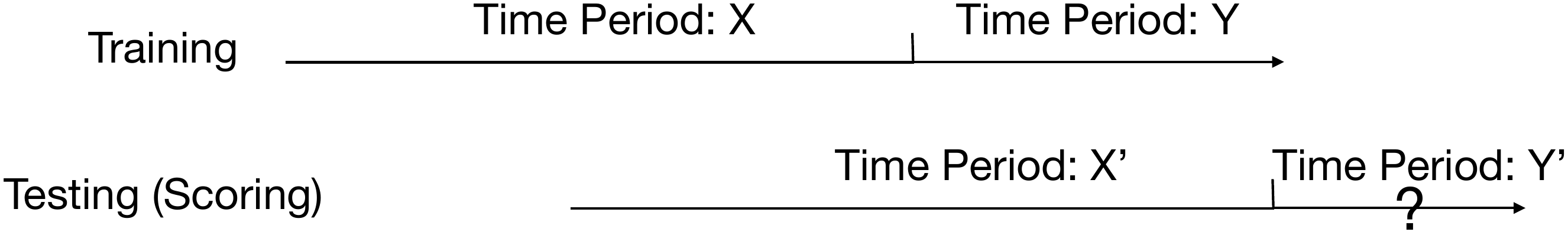}
	\caption{Illustration of training and scoring data splitting. }%
	\label{fig:setup}
\end{wrapfigure}

\subsection{Evaluation methodology and metrics} To evaluate the proposed model, we conduct a series of offline experiments on real-world Amazon Prime Video data. As shown in Figure \ref{fig:setup}, during the training process, we use the watch history of users in a two-year time period ($X$) to predict their watch behaviors in the following two-week period ($Y$). Then, in the scoring (testing) mode, given the watch history in the two-year period $X'$, it can calculate the preference scores and predict users' watch behavior for the following one-week time period $Y'$.  To avoid data leakage, there is no overlap between $Y'$ and ($X$ $\cup$ $Y$).

For each user, we calculate preference scores for a list of candidate movies or TV shows. We rank these titles based on the predicted scores and select the Top-K titles for different categories (i.e.,movies and series). We pick $K=6$ to simulate the use cases in Amazon Prime Video. With the top-k titles and the ground-truth, we adopt 4 different metrics to evaluate the recommendation performance. (1) \textbf{Precision@K} represents the ratio of the actual watched items among the Top-K Recommendation; (2) \textbf{Recall@K} is the ratio of the actual watched items which were uncovered by the recommendations; (3) \textbf{Coverage@K} shows the number of unique items that were recommended for all users, and (4) \textbf{Converted Coverage@K} to indicate the number of unique items recommended to all users that were actually watched.

\smallskip
\noindent\textbf{Data.} We collect and split the data according to the scheme explained in Figure  \ref{fig:setup}. Specifically, the training data with density\footnote{density = $ \frac{\sharp watches}{\sharp users * \sharp items}$} of 0.00672\% has the length of 2-year in total (i.e., $X+Y=2$ years), in which the data for the last 14-day is treated as labels. For the test set, we collect user streaming history in 2-year as input features to predict what user is going to watch in the next 7-day (Time Period $Y'$). 
To evaluate how the proposed model works for cold-start title recommendation we did a series of offline experiments. Here we have a set of items that have no watch history in time period $X$, $Y$ or $X'$ but have at least one watch in time period $Y'$. There are 360 movies and 75 TV series in total in this category. In the following experiments, during the scoring/evaluation, we only calculate preference scores for this set of cold-start titles and rank among them. 

\begin{table}
	\centering
	\scalebox{0.80}{\resizebox{\textwidth}{!}{%
			\begin{tabular}{|l|cc|cc|cc|cc|}
				\hline
				& \multicolumn{2}{l|}{P@6 (\%)} & \multicolumn{2}{c|}{R@6 (\%)}   & \multicolumn{2}{c|}{Cov@6} & \multicolumn{2}{c|}{ConCov@6} \\ \hline
				Model & Movie    &  TV      & Movie   & TV     &Movie   & TV      & Movie   & TV      \\ \hline\hline
				{\it Synopsis} &   (B)  & (B) & (B)  & (B)  & (B) & (B)  &(B) &(B) \\			
				{\it Cover art} & 0.61 &  0.77 & 2.07 &  1.41 &  -30.87 & -26.65 & -25.60 & -23.59  \\ 
				{\it Categorical} &  11.38 & -25.38 &   10.27& -11.76   & 19.24 & -55.50  & 30.13 &-46.38 \\
				{\it NCF-extention}               &  15.38 & 7.69 &   12.75& 8.24   & 44.05 & 4.52  & 70.72 & 15.82 \\
				{\it Con (w/o ID)}  & 13.53 &  1.54 & 12.76 &  1.88 &  33.83 & 0.31 & 36.78 & 4.29  \\
				{\it Att (w/o ID)}  & 17.23 &  15.38 & 14.94 &  16.24 &  65.06& 9..03 & 79.49& 18.49  \\\hline
				{\it Con (w/ ID)} &   17.85 & 18.46 & 14.95  & 19.05  & 79.92& 25.14  &99.15 &39.14 \\
				{\it Att (w/ ID)} &   {\bf 26.76}  & {\bf 25.38} & {\bf 21.88} &  {\bf 24.71}  &  {\bf 167.57} &   {\bf 50.63} & {\bf 176.23} &  {\bf 81.76} \\\hline
			\end{tabular}%
	}}
	\caption{Comparison on warm-start video recommendation. For fair comparison, {\it NCF-extention} extends NCF \cite{he2017neural} by concatenating both the watch history embedding and user-level features embedding with the user ID embedding. (B) indicates the basline for percentage calculation. All numbers are reported in percentage (\%) lift w.r.t. the baseline. {\it Synopsis}, {\it Cover art} and  {\it Categorical} denote the models use only categorical features, synopsis features or cover art features; {\it Con} concatenate all types of metadata directly, and feeds the concatenation into the 1-NN; {\it Att} means to fuse different types of metadata using attention.}
	\label{table:warm-com}
\end{table}

\smallskip
\noindent\textbf{Parameter Setup.} We use positive samples from a user's watch history, and negative samples from randomly sampling a set of items the user hasn't watched before. We tried different negative sampling rates ranging from 1 to 30. We found that more negative samples can lead to better performance. Empirically, the recommendation performance becomes stable when the negative sampling rate is larger than 20. The embedding size before the dot product is also set to be 512. We used Adam optimizer with the learning rate of 0.001.


\subsection{Results}

\subsubsection{Warm-start Video Recommendation Task.} To fully understand how different metadata and the proposed attention layer work for video recommendation, we evaluate the recommendations for the {\it warm-start videos}. The results are summarized in Table \ref{table:warm-com}. First, we compare the models using only one type of metadata. We can see that the categorical features can result in the best offline performance under different evaluation metrics, indicating that the categorical features are more informative in characterizing the warm-start videos compared with other metadata. When we fuse the aforementioned metadata, the proposed attention layer can bring in significant improvement compared to the models using one of the metadata, while the simple concatenation only works slightly better than those models. This observation illustrates the effectiveness of the attention layer. Furthermore, if we take the ID embedding into consideration, under the warm-start setup, the models can perform better than all the models without ID embedding. The reason is that for items with abundant watch history, the ID embedding layer can learn an informative representation, which hints at the necessity of ID embedding in the warm-start setup.


\begin{table}[b]
\centering
\scalebox{0.80}{\resizebox{\textwidth}{!}{%
\begin{tabular}{|l|cc|cc|cc|cc|}
\hline

 & \multicolumn{2}{l|}{P@6 (\%)} & \multicolumn{2}{c|}{R@6 (\%)}   & \multicolumn{2}{c|}{Cov@6} & \multicolumn{2}{c|}{ConCov@6} \\ \hline
 Model & Movie    & TV      & Movie   & TV      &Movie   & TV       & Movie   & TV        \\ \hline\hline
{\it Random}&  (B) & (B) &   (B) & (B)   & {\bf (B)} & {\bf (B)}  & (B) &(B) \\
{\it Att (w/o ID)} &   {\bf 2556.52}  & {\bf 385.42} & {\bf 2976.51}  & {\bf 400.11}  & -26.46 &-9.33  &354.54 &3.85 \\
 {\it Att (w/ ID) }  & 1221.73 &  368.75& 1277.35 &  329.92&  -2.78& {\bf 0.00} & {\bf 486.36} & {\bf 57.69}  \\ \hline
\end{tabular}%
}}
\caption{Comparison on cold-start video recommendation. (B) indicates the baseline for percentage calculation. All numbers are reported in percentage (\%) lift w.r.t. the baseline. {\it Random} is the random recommendation strategy; {\it Att w/o ID} is the two-tower model with Attention layer fusing all types of metadata; 
	 {\bf{\it Att with ID}} is the two-tower model considering all types of metadata and ID embeddings, using Attention layer to fuse those metadata and ID embeddings.}
\label{table:cold-com}
\end{table}

\subsubsection{Cold-start Item Recommendation Task.} We compare the proposed two-tower model with the random baseline in which we just randomly select 6 cold-start titles to each user. For cold-start titles, during evaluation/scoring phase, we set their ID embedding to zero. To examine how the ID embedding will influence the cold-start recommendation, we also compare the two-tower model with all the metadata and the version with metadata and ID embedding. The results are summarized in Table \ref{table:cold-com}. We can see that, both the versions with or without ID embedding can beat the random method significantly, explaining the effectiveness of the two-tower model with metadata. 
Further, the model with ID embedding falls behind the model without ID embedding in terms of precision@6 and recall@6. The reason is that higher weights are usually assigned to the ID embedding by the attention layer in the training phases, and leading to a weak representation for the metadata components. 
Thus if we remove the ID embedding, we can see the improvements for the cold-start item prediction task. Note, the convertedCoverage@6 and Coverage@6 are slightly higher for the two-tower model with ID embedding. We need to carefully consider this tradeoff scenario while designing an appropriate model for cold-start video recommendation.



\subsection{Visualization}
\begin{figure}
	\centering
	\includegraphics[width=0.8\textwidth]{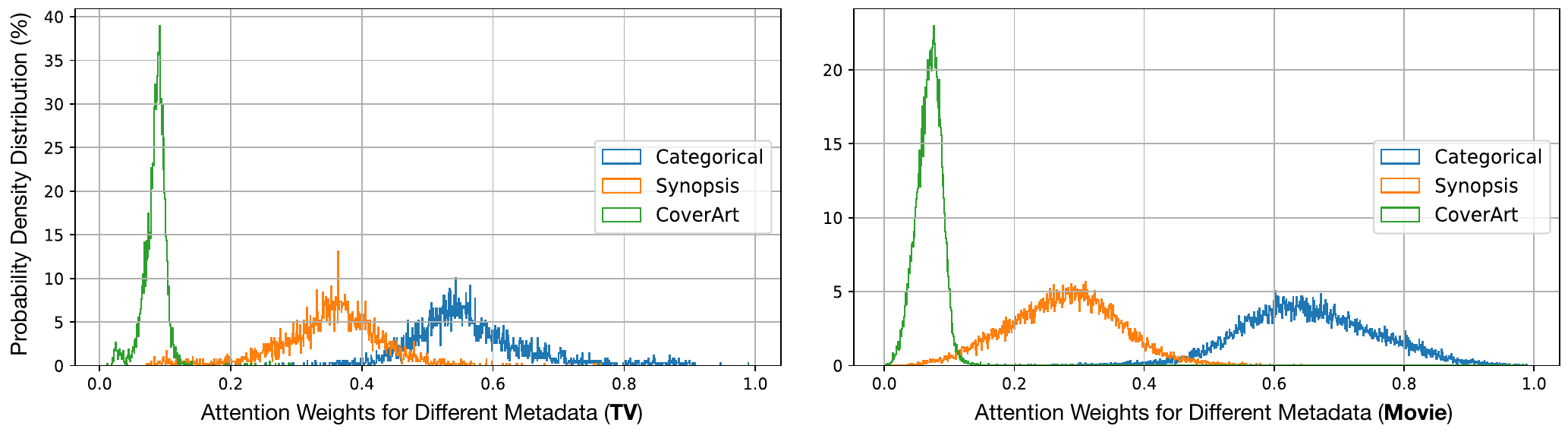}
	\caption{The distribution of attention weights.}%
	\label{fig:weight_category}%
\end{figure}

To examine how attention layer works in controlling the fusion process,  in Figure \ref{fig:weight_category}, we summarize the resulted weights for different types of metadata in TV shows and movies. The results for movie and TV shows are shown individually. We find that for both title categories, cover art feature gets the lowest weights, and categorical features get the highest weights. This is intuitive, since categorical data is rich with important features, whereas the cover art is only represented by a pre-trained embedding, and not fine-tuned for this task. When comparing TV shows and movies, we find that synopsis features are more important for TV shows compared to movies. 

\smallskip

\section{Conclusion}
We proposed a two-tower model for cold- and warm-start item recommendation with heterogeneous metadata. We also explored different types of metadata, including categorical features, cover-art images and synopsis, 
With offline experiments, we show that the proposed framework can produce best recommendations by fusing different types of metadata using attention. By introducing the metadata with the well-designed attention layer, the two-tower model enables the recommendation for cold-start videos and also improve the recommendation for warm-start videos. 

\bibliographystyle{ACM-Reference-Format}
\bibliography{sample-bibliography} 


\begin{thebibliography}{22}


\ifx \showCODEN    \undefined \def \showCODEN     #1{\unskip}     \fi
\ifx \showDOI      \undefined \def \showDOI       #1{#1}\fi
\ifx \showISBNx    \undefined \def \showISBNx     #1{\unskip}     \fi
\ifx \showISBNxiii \undefined \def \showISBNxiii  #1{\unskip}     \fi
\ifx \showISSN     \undefined \def \showISSN      #1{\unskip}     \fi
\ifx \showLCCN     \undefined \def \showLCCN      #1{\unskip}     \fi
\ifx \shownote     \undefined \def \shownote      #1{#1}          \fi
\ifx \showarticletitle \undefined \def \showarticletitle #1{#1}   \fi
\ifx \showURL      \undefined \def \showURL       {\relax}        \fi
\providecommand\bibfield[2]{#2}
\providecommand\bibinfo[2]{#2}
\providecommand\natexlab[1]{#1}
\providecommand\showeprint[2][]{arXiv:#2}

\bibitem[\protect\citeauthoryear{Adomavicius, Sankaranarayanan, Sen, and
  Tuzhilin}{Adomavicius et~al\mbox{.}}{2005}]%
        {adomavicius2005incorporating}
\bibfield{author}{\bibinfo{person}{Gediminas Adomavicius},
  \bibinfo{person}{Ramesh Sankaranarayanan}, \bibinfo{person}{Shahana Sen},
  {and} \bibinfo{person}{Alexander Tuzhilin}.} \bibinfo{year}{2005}\natexlab{}.
\newblock \showarticletitle{Incorporating contextual information in recommender
  systems using a multidimensional approach}.
\newblock \bibinfo{journal}{{\em ACM Transactions on Information systems
  (TOIS)\/}} \bibinfo{volume}{23}, \bibinfo{number}{1} (\bibinfo{year}{2005}),
  \bibinfo{pages}{103--145}.
\newblock


\bibitem[\protect\citeauthoryear{Covington, Adams, and Sargin}{Covington
  et~al\mbox{.}}{2016}]%
        {covington2016deep}
\bibfield{author}{\bibinfo{person}{Paul Covington}, \bibinfo{person}{Jay
  Adams}, {and} \bibinfo{person}{Emre Sargin}.}
  \bibinfo{year}{2016}\natexlab{}.
\newblock \showarticletitle{Deep neural networks for youtube recommendations}.
  In \bibinfo{booktitle}{{\em Proceedings of the 10th ACM conference on
  recommender systems}}. \bibinfo{pages}{191--198}.
\newblock


\bibitem[\protect\citeauthoryear{Davidson, Liebald, Liu, Nandy, Van~Vleet,
  Gargi, Gupta, He, Lambert, Livingston, et~al\mbox{.}}{Davidson
  et~al\mbox{.}}{2010}]%
        {davidson2010youtube}
\bibfield{author}{\bibinfo{person}{James Davidson}, \bibinfo{person}{Benjamin
  Liebald}, \bibinfo{person}{Junning Liu}, \bibinfo{person}{Palash Nandy},
  \bibinfo{person}{Taylor Van~Vleet}, \bibinfo{person}{Ullas Gargi},
  \bibinfo{person}{Sujoy Gupta}, \bibinfo{person}{Yu He}, \bibinfo{person}{Mike
  Lambert}, \bibinfo{person}{Blake Livingston}, {et~al\mbox{.}}}
  \bibinfo{year}{2010}\natexlab{}.
\newblock \showarticletitle{The YouTube video recommendation system}. In
  \bibinfo{booktitle}{{\em Proceedings of the fourth ACM conference on
  Recommender systems}}. \bibinfo{pages}{293--296}.
\newblock


\bibitem[\protect\citeauthoryear{Elkahky, Song, and He}{Elkahky
  et~al\mbox{.}}{2015}]%
        {elkahky2015multi}
\bibfield{author}{\bibinfo{person}{Ali~Mamdouh Elkahky}, \bibinfo{person}{Yang
  Song}, {and} \bibinfo{person}{Xiaodong He}.} \bibinfo{year}{2015}\natexlab{}.
\newblock \showarticletitle{A multi-view deep learning approach for cross
  domain user modeling in recommendation systems}. In \bibinfo{booktitle}{{\em
  Proceedings of the 24th international conference on world wide web}}.
  \bibinfo{pages}{278--288}.
\newblock


\bibitem[\protect\citeauthoryear{Gantner, Drumond, Freudenthaler, Rendle, and
  Schmidt-Thieme}{Gantner et~al\mbox{.}}{2010}]%
        {gantner2010learning}
\bibfield{author}{\bibinfo{person}{Zeno Gantner}, \bibinfo{person}{Lucas
  Drumond}, \bibinfo{person}{Christoph Freudenthaler}, \bibinfo{person}{Steffen
  Rendle}, {and} \bibinfo{person}{Lars Schmidt-Thieme}.}
  \bibinfo{year}{2010}\natexlab{}.
\newblock \showarticletitle{Learning attribute-to-feature mappings for
  cold-start recommendations}. In \bibinfo{booktitle}{{\em 2010 IEEE
  International Conference on Data Mining}}. IEEE, \bibinfo{pages}{176--185}.
\newblock


\bibitem[\protect\citeauthoryear{Gong, Kaya, and Tintarev}{Gong
  et~al\mbox{.}}{2020}]%
        {gongcontextual}
\bibfield{author}{\bibinfo{person}{Boning Gong}, \bibinfo{person}{Mesut Kaya},
  {and} \bibinfo{person}{Nava Tintarev}.} \bibinfo{year}{2020}\natexlab{}.
\newblock \showarticletitle{Contextual Personalized Re-Ranking of Music
  Recommendations through Audio Features}. In \bibinfo{booktitle}{{\em CARS
  Workshop at ACM RecSys}}.
\newblock


\bibitem[\protect\citeauthoryear{He, Zhang, Ren, and Sun}{He
  et~al\mbox{.}}{2016}]%
        {he2016deep}
\bibfield{author}{\bibinfo{person}{Kaiming He}, \bibinfo{person}{Xiangyu
  Zhang}, \bibinfo{person}{Shaoqing Ren}, {and} \bibinfo{person}{Jian Sun}.}
  \bibinfo{year}{2016}\natexlab{}.
\newblock \showarticletitle{Deep residual learning for image recognition}. In
  \bibinfo{booktitle}{{\em Proceedings of the IEEE conference on computer
  vision and pattern recognition}}. \bibinfo{pages}{770--778}.
\newblock


\bibitem[\protect\citeauthoryear{He, Liao, Zhang, Nie, Hu, and Chua}{He
  et~al\mbox{.}}{2017}]%
        {he2017neural}
\bibfield{author}{\bibinfo{person}{Xiangnan He}, \bibinfo{person}{Lizi Liao},
  \bibinfo{person}{Hanwang Zhang}, \bibinfo{person}{Liqiang Nie},
  \bibinfo{person}{Xia Hu}, {and} \bibinfo{person}{Tat-Seng Chua}.}
  \bibinfo{year}{2017}\natexlab{}.
\newblock \showarticletitle{Neural collaborative filtering}. In
  \bibinfo{booktitle}{{\em Proceedings of the 26th international conference on
  world wide web}}. \bibinfo{pages}{173--182}.
\newblock


\bibitem[\protect\citeauthoryear{Huang, He, Gao, Deng, Acero, and Heck}{Huang
  et~al\mbox{.}}{2013}]%
        {huang2013learning}
\bibfield{author}{\bibinfo{person}{Po-Sen Huang}, \bibinfo{person}{Xiaodong
  He}, \bibinfo{person}{Jianfeng Gao}, \bibinfo{person}{Li Deng},
  \bibinfo{person}{Alex Acero}, {and} \bibinfo{person}{Larry Heck}.}
  \bibinfo{year}{2013}\natexlab{}.
\newblock \showarticletitle{Learning deep structured semantic models for web
  search using clickthrough data}. In \bibinfo{booktitle}{{\em Proceedings of
  the 22nd ACM international conference on Information \& Knowledge
  Management}}. \bibinfo{pages}{2333--2338}.
\newblock


\bibitem[\protect\citeauthoryear{Koren, Bell, and Volinsky}{Koren
  et~al\mbox{.}}{2009}]%
        {koren2009matrix}
\bibfield{author}{\bibinfo{person}{Yehuda Koren}, \bibinfo{person}{Robert
  Bell}, {and} \bibinfo{person}{Chris Volinsky}.}
  \bibinfo{year}{2009}\natexlab{}.
\newblock \showarticletitle{Matrix factorization techniques for recommender
  systems}.
\newblock \bibinfo{journal}{{\em Computer\/}} \bibinfo{volume}{42},
  \bibinfo{number}{8} (\bibinfo{year}{2009}), \bibinfo{pages}{30--37}.
\newblock


\bibitem[\protect\citeauthoryear{Lee, Im, Jang, Cho, and Chung}{Lee
  et~al\mbox{.}}{2019}]%
        {lee2019melu}
\bibfield{author}{\bibinfo{person}{Hoyeop Lee}, \bibinfo{person}{Jinbae Im},
  \bibinfo{person}{Seongwon Jang}, \bibinfo{person}{Hyunsouk Cho}, {and}
  \bibinfo{person}{Sehee Chung}.} \bibinfo{year}{2019}\natexlab{}.
\newblock \showarticletitle{Melu: Meta-learned user preference estimator for
  cold-start recommendation}. In \bibinfo{booktitle}{{\em Proceedings of the
  25th ACM SIGKDD International Conference on Knowledge Discovery \& Data
  Mining}}. \bibinfo{pages}{1073--1082}.
\newblock


\bibitem[\protect\citeauthoryear{Lee and Abu-El-Haija}{Lee and
  Abu-El-Haija}{2017}]%
        {lee2017large}
\bibfield{author}{\bibinfo{person}{Joonseok Lee} {and} \bibinfo{person}{Sami
  Abu-El-Haija}.} \bibinfo{year}{2017}\natexlab{}.
\newblock \showarticletitle{Large-scale content-only video recommendation}. In
  \bibinfo{booktitle}{{\em Proceedings of the IEEE International Conference on
  Computer Vision Workshops}}. \bibinfo{pages}{987--995}.
\newblock


\bibitem[\protect\citeauthoryear{Mikolov, Chen, Corrado, and Dean}{Mikolov
  et~al\mbox{.}}{2013a}]%
        {mikolov2013efficient}
\bibfield{author}{\bibinfo{person}{Tomas Mikolov}, \bibinfo{person}{Kai Chen},
  \bibinfo{person}{Greg Corrado}, {and} \bibinfo{person}{Jeffrey Dean}.}
  \bibinfo{year}{2013}\natexlab{a}.
\newblock \showarticletitle{Efficient estimation of word representations in
  vector space}.
\newblock \bibinfo{journal}{{\em arXiv preprint arXiv:1301.3781\/}}
  (\bibinfo{year}{2013}).
\newblock


\bibitem[\protect\citeauthoryear{Mikolov, Sutskever, Chen, Corrado, and
  Dean}{Mikolov et~al\mbox{.}}{2013b}]%
        {mikolov2013distributed}
\bibfield{author}{\bibinfo{person}{Tomas Mikolov}, \bibinfo{person}{Ilya
  Sutskever}, \bibinfo{person}{Kai Chen}, \bibinfo{person}{Greg~S Corrado},
  {and} \bibinfo{person}{Jeff Dean}.} \bibinfo{year}{2013}\natexlab{b}.
\newblock \showarticletitle{Distributed representations of words and phrases
  and their compositionality}. In \bibinfo{booktitle}{{\em Advances in neural
  information processing systems}}. \bibinfo{pages}{3111--3119}.
\newblock


\bibitem[\protect\citeauthoryear{Rybakov, Mohan, Misra, LeGrand, Joseph, Chung,
  Singh, You, Nalisnick, Dirac, and Luo}{Rybakov et~al\mbox{.}}{2018}]%
        {Rybakov2018TheEO}
\bibfield{author}{\bibinfo{person}{O. Rybakov}, \bibinfo{person}{V. Mohan},
  \bibinfo{person}{A. Misra}, \bibinfo{person}{S. LeGrand}, \bibinfo{person}{R.
  Joseph}, \bibinfo{person}{Kiuk Chung}, \bibinfo{person}{Siddharth Singh},
  \bibinfo{person}{Q. You}, \bibinfo{person}{Eric~T. Nalisnick},
  \bibinfo{person}{Leo Dirac}, {and} \bibinfo{person}{Runfei Luo}.}
  \bibinfo{year}{2018}\natexlab{}.
\newblock \showarticletitle{The Effectiveness of a two-Layer Neural Network for
  Recommendations}. In \bibinfo{booktitle}{{\em ICLR Workshop}}.
\newblock


\bibitem[\protect\citeauthoryear{Vaswani, Shazeer, Parmar, Uszkoreit, Jones,
  Gomez, Kaiser, and Polosukhin}{Vaswani et~al\mbox{.}}{2017}]%
        {vaswani2017attention}
\bibfield{author}{\bibinfo{person}{Ashish Vaswani}, \bibinfo{person}{Noam
  Shazeer}, \bibinfo{person}{Niki Parmar}, \bibinfo{person}{Jakob Uszkoreit},
  \bibinfo{person}{Llion Jones}, \bibinfo{person}{Aidan~N Gomez},
  \bibinfo{person}{{\L}ukasz Kaiser}, {and} \bibinfo{person}{Illia
  Polosukhin}.} \bibinfo{year}{2017}\natexlab{}.
\newblock \showarticletitle{Attention is all you need}. In
  \bibinfo{booktitle}{{\em Advances in neural information processing systems}}.
  \bibinfo{pages}{5998--6008}.
\newblock


\bibitem[\protect\citeauthoryear{Volkovs, Yu, and Poutanen}{Volkovs
  et~al\mbox{.}}{2017}]%
        {volkovs2017dropoutnet}
\bibfield{author}{\bibinfo{person}{Maksims Volkovs}, \bibinfo{person}{Guang~Wei
  Yu}, {and} \bibinfo{person}{Tomi Poutanen}.} \bibinfo{year}{2017}\natexlab{}.
\newblock \showarticletitle{DropoutNet: Addressing Cold Start in Recommender
  Systems.}. In \bibinfo{booktitle}{{\em NIPS}}. \bibinfo{pages}{4957--4966}.
\newblock


\bibitem[\protect\citeauthoryear{Wang, Ding, and Caverlee}{Wang
  et~al\mbox{.}}{2021}]%
        {wang2021sequential}
\bibfield{author}{\bibinfo{person}{Jianling Wang}, \bibinfo{person}{Kaize
  Ding}, {and} \bibinfo{person}{James Caverlee}.}
  \bibinfo{year}{2021}\natexlab{}.
\newblock \showarticletitle{Sequential Recommendation for Cold-start Users with
  Meta Transitional Learning}. In \bibinfo{booktitle}{{\em Proceedings of the
  44th International ACM SIGIR Conference on Research and Development in
  Information Retrieval}}. \bibinfo{pages}{1783--1787}.
\newblock


\bibitem[\protect\citeauthoryear{Wang, Ding, Hong, Liu, and Caverlee}{Wang
  et~al\mbox{.}}{2020a}]%
        {wang2020next}
\bibfield{author}{\bibinfo{person}{Jianling Wang}, \bibinfo{person}{Kaize
  Ding}, \bibinfo{person}{Liangjie Hong}, \bibinfo{person}{Huan Liu}, {and}
  \bibinfo{person}{James Caverlee}.} \bibinfo{year}{2020}\natexlab{a}.
\newblock \showarticletitle{Next-item recommendation with sequential
  hypergraphs}. In \bibinfo{booktitle}{{\em Proceedings of the 43rd
  international ACM SIGIR conference on research and development in information
  retrieval}}. \bibinfo{pages}{1101--1110}.
\newblock


\bibitem[\protect\citeauthoryear{Wang, Zhu, and Caverlee}{Wang
  et~al\mbox{.}}{2020b}]%
        {wang2020user}
\bibfield{author}{\bibinfo{person}{Jianling Wang}, \bibinfo{person}{Ziwei Zhu},
  {and} \bibinfo{person}{James Caverlee}.} \bibinfo{year}{2020}\natexlab{b}.
\newblock \showarticletitle{User Recommendation in Content Curation Platforms}.
  In \bibinfo{booktitle}{{\em Proceedings of the 13th International Conference
  on Web Search and Data Mining}}. \bibinfo{pages}{627--635}.
\newblock


\bibitem[\protect\citeauthoryear{Wang, Zhao, Yi, Yang, Cheng, Hong, Tjoa, Kang,
  Ettinger, and Chi}{Wang et~al\mbox{.}}{2019}]%
        {wangimproving}
\bibfield{author}{\bibinfo{person}{Ruoxi Wang}, \bibinfo{person}{Zhe Zhao},
  \bibinfo{person}{Xinyang Yi}, \bibinfo{person}{Ji Yang},
  \bibinfo{person}{Derek~Zhiyuan Cheng}, \bibinfo{person}{Lichan Hong},
  \bibinfo{person}{Steve Tjoa}, \bibinfo{person}{Jieqi Kang},
  \bibinfo{person}{Evan Ettinger}, {and} \bibinfo{person}{H Chi}.}
  \bibinfo{year}{2019}\natexlab{}.
\newblock \showarticletitle{Improving Relevance Prediction with Transfer
  Learning in Large-scale Retrieval Systems}. In \bibinfo{booktitle}{{\em
  Proceedings of the 1st Adaptive \& Multitask Learning Workshop}}.
\newblock


\bibitem[\protect\citeauthoryear{Yang, Yi, Zhiyuan~Cheng, Hong, Li,
  Xiaoming~Wang, Xu, and Chi}{Yang et~al\mbox{.}}{2020}]%
        {yang2020mixed}
\bibfield{author}{\bibinfo{person}{Ji Yang}, \bibinfo{person}{Xinyang Yi},
  \bibinfo{person}{Derek Zhiyuan~Cheng}, \bibinfo{person}{Lichan Hong},
  \bibinfo{person}{Yang Li}, \bibinfo{person}{Simon Xiaoming~Wang},
  \bibinfo{person}{Taibai Xu}, {and} \bibinfo{person}{Ed~H Chi}.}
  \bibinfo{year}{2020}\natexlab{}.
\newblock \showarticletitle{Mixed negative sampling for learning two-tower
  neural networks in recommendations}. In \bibinfo{booktitle}{{\em Companion
  Proceedings of the Web Conference 2020}}. \bibinfo{pages}{441--447}.
\newblock


\end{thebibliography}

\end{document}